\documentclass[aps,prb,reprint,superscriptaddress,longbibliography]{revtex4-2}

\usepackage[utf8]{inputenc}
\usepackage[T1]{fontenc}

\usepackage{amsmath,amssymb,bm}
\usepackage{graphicx}
\usepackage{dcolumn}               % align on decimal points in tables
\usepackage[caption=false]{subfig} % subfigures compatible with REVTeX

\usepackage{tikz}
\usetikzlibrary{decorations.text,calc,decorations.pathmorphing}
\tikzset{
  vertex/.style = {font=\sffamily\bfseries, text=white, shape=circle, ball color=black, minimum size=1mm,},
  bridge/.style = {thick, draw=black, bend right=20},
  number/.style = {font=\sffamily\bfseries, text=white, draw, fill=red, circle, minimum size=6mm},
  threej/.style={fill=black, circle, minimum size=0.5mm, scale=0.5}
}

\usepackage[colorlinks=true,allcolors=blue]{hyperref}

\begin{document}

\title{Multichannel active-space embedding of atomic multiplets in plane-wave DFT/PAW for core-level spectroscopies}

\author{Alessandro Mirone}\email{mirone@esrf.fr}
\affiliation{European Synchrotron Radiation Facility, 71, Avenue des Martyrs, Grenoble F-38000, France}
\author{Mauro Rovezzi}
\affiliation{Univ. Grenoble Alpes, CNRS, IRD, Irstea, Météo France, OSUG, FAME, 38000 Grenoble, France}
\author{Christoph Sahle}
\affiliation{European Synchrotron Radiation Facility, 71, Avenue des Martyrs, Grenoble F-38000, France}
\author{Alessandro Longo}
\affiliation{Univ. Grenoble Alpes, CNRS, IRD, Irstea, Météo France, OSUG, FAME, 38000 Grenoble, France}
\affiliation{Istituto per lo Studio dei Materiali Nanostrutturati, Consiglio Nazionale delle Ricerche, Via Ugo La Malfa 153, Palermo, 90146 Italy}

\begin{abstract}
We introduce an active-space embedding framework for core-level spectroscopies that connects localized atomic multiplets to continuum resonances within a plane-wave DFT/PAW description. The approach is complementary to widely used core-level Bethe--Salpeter implementations based on a two-particle (core-exciton) picture with typically static screening: here a correlated multiplet manifold of the absorber (including the core hole and open-shell configurations) is coherently coupled to a plane-wave photoelectron, enabling a unified treatment of localized multiplet structure and continuum lineshapes. Spectra are computed in a general time-domain formulation equivalent to Fermi's golden rule: a transition operator tailored to the specific spectroscopy technique is applied to the correlated ground state to generate an excited wavepacket, and the corresponding wavepacket autocorrelation function is evaluated without explicit real-time propagation, using Lanczos tridiagonalization or the kernel polynomial method; the spectral intensity follows from its Fourier representation. We validate the method at the Ce \(N_{4,5}\) edges, reproducing in quantitative agreement with experiment both high-\(Q\) multiplet features and the low-\(Q\) giant dipole resonance continuum, including a characteristic low-energy shoulder relevant for robust Ce valence assignments. The implementation is available open-source within the Quantum ESPRESSO XSPECTRA package (xspectruplet mode), together with reproducible inputs and scripts.
\end{abstract}

\maketitle

% ============================================================
\section{Introduction}
% ============================================================

Since its original formulation, density functional theory (DFT)~\cite{Kohn} has evolved into a versatile tool that accurately describes ground-state properties of complex, many-electron systems. By employing advanced exchange–correlation functionals—such as hybrid and meta-GGA forms—DFT now often yields results that closely track experimental findings, even for systems with pronounced electron–electron correlations~\cite{pbe,chemreview}. In parallel, many-body approaches, including the GW approximation~\cite{gw} and dynamical mean-field theory (DMFT)~\cite{dmft}, capture correlation effects and electron dynamics with remarkable fidelity.

However, when the goal shifts from ground states to spectroscopy, the situation changes dramatically. X-ray spectroscopies probe states far from the ground state; core excitations produce a core hole and a photoelectron whose entanglement with open-shell electrons yields intricate multiplet structures and resonances with strong continuum character.

For \emph{localized} excitations, multiplet approaches remain remarkably predictive. Early implementations based on small active spaces and symmetry exploitation~\cite{tt} evolved into charge-transfer multiplet theory~\cite{deGroot2008}, arbitrary symmetry~\cite{amarcord}, and \emph{ab initio} parameter extraction via Wannier Hamiltonians~\cite{quanty,vanadiohilb,w90}. Yet, the Hilbert space of the active spin–orbitals grows exponentially, limiting the explicit treatment of continuum states.

Alternatively, one may start from \emph{ab initio} single-particle frameworks and add correlations: self-energy corrections to the single-particle propagator (GW), or corrections in the particle--hole sector via the Bethe--Salpeter equation (BSE)~\cite{Vinson2011,Prendergast2006,Vorwerk2017}. While the BSE provides, in principle, an exact representation of linear response, practical core-level implementations typically rely on approximate (often static) kernels and on a restricted description of the absorber degrees of freedom. Implementations such as OCEAN~\cite{Vinson2011,keith} model a photoexcited electron coupled to a localized core hole, yielding a practical core-exciton treatment. For open-shell ions (e.g., Ce$^{3+}$ with $4f^1$), however, the relevant final states span a correlated manifold of ionic configurations coupled to the photoelectron, which is inefficient to represent within the standard two-particle core-exciton picture used in most practical BSE-based calculations.
From a formal many-body perspective, the linear density response can be obtained from the four-point polarizability (two-particle Green's function) via the Bethe--Salpeter equation (BSE), which is formally exact.~\cite{Onida2002RMP}
In practice, however, most core-level implementations rely on a two-particle core-exciton reduction with statically screened interactions and additional approximations, which can hinder the description of dynamical correlation effects (e.g., satellite structures and higher-order excitations) and of strongly correlated open-shell manifolds coupled to the continuum.~\cite{CudazzoReining2020PRR}
Frequency-dependent (dynamical) kernels have been explored to recover such effects (including double excitations), at the price of increased complexity and, in approximate schemes, the possible appearance of spurious solutions.~\cite{AuthierLoos2020JCP}

Here we take a complementary route tailored to open-shell absorbers, going beyond the standard static-kernel two-particle core-exciton practice by explicitly embedding a correlated atomic active space and coherently coupling it to the plane-wave/PAW photoelectron framework of XSPECTRA~\cite{bunau,mauri1,mauri2} within Quantum ESPRESSO~\cite{QE-2017,QE-2009}.

Projector augmented‑wave (PAW) pseudopotentials~\cite{Blochl1994,KresseJoubert1999} are pivotal: besides accurate pseudopotentials, PAW projectors provide localized channels to map excited atomic states onto the plane‑wave pseudobasis. Unlike the original XSPECTRA—which solves a single‑particle problem—we represent the photoelectron as a superposition of states, each given by the tensor product of a single‑particle plane‑wave function and an ionic basis vector $|a_i\rangle$.
Here we discuss how this approach,  named {\it xspectruplet-mode} within the new XSPECTRA code, captures both localized multiplet peaks and the continuum component of the giant dipole resonance (GDR), and we illustrate it on the Ce $N_{4,5}$ edges.
The term giant dipole resonance originates from photonuclear absorption studies \cite{BaldwinKlaiber1947,LevingerBethe1950}; in rare-earth core spectroscopies it is used by analogy to denote a strong, broad dipolar resonance with substantial continuum character.

% Indistinguishability is preserved and Pauli exclusion enforced via projectors, while ionic Hamiltonians imported from a multiplet code scatter amplitudes across replicated photoelectron channels tied to different ionic configurations. 
\paragraph*{Intuitive picture}
Before developing the full formalism, it is useful to outline, as an illustration of the importance of entanglement,  the exchange‑driven mechanism underlying the GDR (Fig.~\ref{fig:birdeye}). At the Ce $N_{4,5}$ edges, a photon with multipolarity $(l_{\mathrm{ph}},m_{\mathrm{ph}})$ promotes a $4d$ electron $(l_d,m)$ to $4f$ $(l_f,m{+}m_{\mathrm{ph}})$, creating a coherent superposition of core–photoelectron pairs (Fig.~\ref{fig:birdeye}, left). The first‑order exchange contribution to the resonance energy can be cast as the trace of the product between the excited‑state density matrix and an exchange vertex of rank $K$ (Fig.~\ref{fig:birdeye}, right). By orthogonality and completeness of Wigner $3j$ symbols~\cite{Yutsis,gordon}, the dominant contribution arises for $K=l_{\mathrm{ph}}$; other $K$ are orthogonal and do not contribute. Consequently, dipole excitations ($l_{\mathrm{ph}}{=}1$) experience a large shift (of order $2G_1$), whereas higher multipoles are much weaker (e.g., $\sim 0.38\,G_3$ and $\sim 0.28\,G_5$), with $G_K$ the $4d$–$4f$ exchange Slater integral of rank $K$. The hierarchy stems from the structure of the superposition: for $l_{\mathrm{ph}}{=}1$ many core–electron pairs interfere constructively, whereas for $l_{\mathrm{ph}}{=}5$ the allowed channels are severely constrained (e.g., for $m_{\mathrm{ph}}{=}5$ only $m{=}{-}2$ is promoted).
Crucially, the exchange vertex is a short-range, near-core interaction: it acts only where the photoelectron has finite amplitude inside the absorber and overlaps the highly localized $4d$ hole. In the dipole channel this contribution is repulsive and large—of order $2G_1 \!\approx\! 30$\,eV for Ce—so it lifts the would-be localized $4d^{-1}4f^{1}$ configuration above the ionization threshold, endowing it with strong continuum character and producing a GDR. In other words, the exchange term can “jump” the barrier between localized states and the continuum. The effect is also \emph{resonance-selective}: it is strong for the dipole ($K{=}1$) channel and much weaker for higher multipoles ($K{=}3,5$), so the amount of spectral-weight transfer into the continuum depends on the specific resonance. In our formulation this near-core exchange arises naturally as an interference effect between the entangled components; the rigorous development follows in Sec.~\ref{sec:theory}.

\begin{figure}[t]
  \centering
  % Left-align the whole multi-panel block and optionally shift it left
  \makebox[\linewidth][l]{%
    % Optional global left shift (tune or remove if not needed)
    \hspace*{-0.6cm}%
    % --- Left block: panels (a) and (b) ---
    \hspace{0.3cm}
    \begin{minipage}[t]{0.45\linewidth}
      \centering
      \begin{tikzpicture}[>=stealth, scale=1]
        \node[fill=black, circle, minimum size=0.5mm, scale=0.5 ] (b) at (-12,0) {};
        \node[] (b1) at (-10.5,-0.5) {};
        \node[] (b2) at (-10.5,0.5) {};
        \coordinate (ab) at (-13,0);
        \node[above=0.9cm, anchor=east] at (b) {{\bf a)} Excited State};
        \draw[->, bend left=20, line width=0.5mm] (b) to node[pos=0.6, below] {$l_{f}\, m_f $} (b2);
        \draw[->, bend left=20, line width=0.5mm] (b1) to node[pos=0.6, below] {$l_{d}\, m_d$} (b);
        \draw[decorate, decoration={snake}] (ab) -> node[left, below] {$l_{ph}\, m_{ph}$} (b);
        \node[] (c2) at (-12.2,-2.3) {};
        \node[] (v1) at ($(c2) +(0,0)$) {};
        \node[] (v2) at ($(c2) +(0.8,0)$) {};
        \node[above=1cm, anchor=east] at (v2) {{\bf b)} Exchange Vertex};
        \node[fill=black, circle, minimum size=0.5mm, scale=0.5] at (v1)  {};
        \node[fill=black, circle, radius=0.5mm, scale=0.5]  at (v2)  {};
        \node[] (va1) at ($(v1)+(-1,-0.5)$) {};
        \node[] (vb1) at ($(v1)+(-1,+0.5)$) {};
        \draw[->, bend left=20, line width=0.5mm] (vb1) to node[pos=0.,below] {$l_{f}\, m_f$}  (v1);
        \draw[->, bend left=20, line width=0.5mm] (v1) to node[pos=1.,above] {$l_{d}\, m_d$}  (va1);
        \draw[decorate, decoration={snake}] (v1) -> node[below] {$K$} (v2);
        \node[] (va2) at ($(v2)+(1,-0.5)$) {};
        \node[] (vb2) at ($(v2)+(1,+0.5)$) {};
        \draw[->, bend left=20, line width=0.5mm] (v2) to node[pos=0.8,below] {$l_{f}\, m^\prime_f$}  (vb2);
        \draw[->, bend left=20, line width=0.5mm] (va2) to node[pos=0.2,below] {$l_{d}\, m^\prime_d$}  (v2);
      \end{tikzpicture}
    \end{minipage}%
    \hspace{0.6cm}
    % --- Right block: panel (c) ---
    \begin{minipage}[t]{0.33\linewidth}
      \centering
      \begin{tikzpicture}[>=stealth, scale=1]
        \node[] (c) at (0,0) {};
        \pgfmathsetmacro{\sh}{1} \pgfmathsetmacro{\sv}{1.5}
        \node[threej] (v1) at ($(c)+(-\sh,+\sv)$) {};
        \node[threej] (v2) at ($(c)+(\sh,+\sv)$) {};
        \node[threej] (v3) at ($(c)+(\sh,-\sv)$) {};
        \node[threej] (v4) at ($(c)+(-\sh,-\sv)$) {};
        \draw[->, bend right=-40, line width=0.5mm] (v4) to node[pos=0.6, left] {$l_{d}$} (v1);
        \draw[->, bend left= 40, line width=0.5mm]  (v1) to node[pos=0.6, right] {$l_{f}$} (v4);
        \draw[ decorate, decoration={snake}] (v4) -> node[below] {$K$} (v3);
        \draw[ decorate, decoration={snake}] (v1) -> node[below] {$l_{ph}$} (v2);
        \draw[->, bend right=-40, line width=0.5mm] (v3) to node[pos=0.6, left] {$l_{d}$} (v2);
        \draw[->, bend left= 40, line width=0.5mm]  (v2) to node[pos=0.6, right] {$l_{f}$} (v3);
        \node[above=1cm, anchor=center, align=center] at ($(v1)!0.5!(v2)$)
          {{\bf c)} Trace of density\\matrix $\times$ exchange};
      \end{tikzpicture}
    \end{minipage}%
  }% end makebox

  \caption{Bird's-eye view of the leading exchange contribution to the dipolar resonance energy at the Ce $N_{4,5}$ edges.
  (a) Schematic excitation step: a photon of multipolarity $(l_{\mathrm{ph}},m_{\mathrm{ph}})$ promotes a $4d$ electron into the $4f$ manifold, creating a coherent superposition of core--photoelectron pairs.
  (b) The exchange vertex.
  (c) The corresponding first-order exchange shift can be written as the trace of the excited-state density matrix times an exchange vertex of rank $K$ (snake line), where solid arrows denote electronic angular-momentum channels.
  In spherical symmetry (and after averaging over $m_{\mathrm{ph}}$), orthogonality of Wigner $3j$ symbols selects $K=l_{\mathrm{ph}}$, strongly enhancing the dipole ($K{=}1$) contribution.}
  \label{fig:birdeye}
\end{figure}

\paragraph*{Practical relevance: Ce $N_{4,5}$ thresholds and valence assignment.}
In recent \emph{in situ} studies on ceria-based systems, the Ce $N_{4,5}$ edge has been widely used as a valence fingerprint to quantify Ce$^{4+}$ vs Ce$^{3+}$ fractions~\cite{Longo2023,Das2024}. While sharp high-$Q$ features are well captured by multiplet models and provide robust constraints on Slater reductions, the low-$Q$ response is dominated by the giant dipole resonance (GDR), whose broad lineshape often presents a low-energy shoulder. If interpreted within a purely localized picture, this shoulder can be mistakenly assigned to a chemically shifted line from reduced species, biasing the inferred Ce$^{3+}$/Ce$^{4+}$ ratio. Our entangled propagation clarifies that this shoulder is a continuum effect—driven by the $K{=}1$ exchange rather than a separate atomic peak.

% ============================================================
\section{Theory}
% ============================================================
\label{sec:theory}
We consider the scattering atom after excitation of the photoelectron, and enumerate by $M$ the number of ionic configurations relevant to the multiplet structure. Each ionic state is a vector $|a_i\rangle$ in a chosen ionic Hilbert basis. The photoelectron is represented by $2M$ spinless plane-wave (PW) channels $|\psi^i\rangle$ (spin is encoded in the index $i$, first $M$ for spin-down, next $M$ for spin-up), and the total \emph{entangled} state is
\begin{equation}
  \left| \widehat{\psi} \right\rangle = \sum_{i=1}^{2M} \left| a_{i} \right\rangle \otimes \left| \psi^{i} \right\rangle,
  \label{eq:initialwf}
\end{equation}
In the following, we refer to a generic vector $\left|\widehat{\psi}\right\rangle$ as an \emph{entangled state}, and to the specific initial condition $\left|\widehat{\psi}_0\right\rangle$ created by the probe as the \emph{(entangled) excited wavepacket}.
\noindent
Throughout this work, the term ``ionic'' refers to localized \emph{electronic} Fock-space configurations of the absorber (core hole and valence/open-shell electrons), not to nuclear/vibrational degrees of freedom.
The meaning of the symbols and the construction of the ionic states are detailed below, together with the 
time evolution operator:
\begin{align}
  \partial_t \left| \widehat{\psi} \right\rangle &= \widehat{D}_t \left| \widehat{\psi} \right\rangle.
  \label{eq:time_evolution}
\end{align}
which defines the formal generator of the evolution operator; its action $e^{t\widehat D_t}$ on the (entangled) excited wavepacket defined below  is evaluated by Lanczos/KPM recursions, without explicit real-time stepping.
Once these two key elements—the wavefunction and the time evolution—are defined,
the spectrum is obtained from the formal time evolution of an \emph{(entangled) excited wavepacket}
$\left| \widehat{\psi}_0 \right\rangle$ generated by applying a probe-specific transition operator
to the correlated ground state:
\begin{equation}
  \left| \widehat{\psi}_0(\mathbf{Q}) \right\rangle \;=\; \widehat{O}(\mathbf{Q}) \left| g \right\rangle \, .
  \label{eq:excited_wavepacket}
\end{equation}
Here $\left| g \right\rangle$ is the correlated ground state and $\widehat{O}(\mathbf{Q})$ is adapted to the specific spectroscopy technique (and carries the dependence on momentum transfer and polarization when relevant).
For example, in nonresonant inelastic x-ray scattering one may take $\widehat{O}(\mathbf{Q})=\sum_j e^{i\mathbf{Q}\cdot\mathbf{r}_j}$, whereas in x-ray absorption $\widehat{O}$ reduces to the appropriate dipole/quadrupole operator.
All selection rules and $Q$-dependence therefore enter through $\widehat{O}$ and the construction of $\left| \widehat{\psi}_0(\mathbf{Q}) \right\rangle$; for notational simplicity we will often omit the explicit $\mathbf{Q}$ label.
In a correlation-function formulation equivalent to Fermi's golden rule, the spectrum can then be written as:
 \begin{equation}
   f(\omega) = \int_{-\infty}^{\infty} 
   \langle \widehat{\psi}_0 | \widehat{S} \, e^{\, t \widehat{D}_t} | \widehat{\psi}_0 \rangle 
   \, e^{i \omega t} \, dt,
   \label{eq:spectra}
 \end{equation}
which can be evaluated either by the Kernel Polynomial Method (KPM)~\cite{kernel} 
or via a continued fraction obtained from Lanczos tridiagonalization~\cite{dagotto}.
Here $\widehat{S}$ denotes the overlap (metric) operator in the entangled space; it reduces to the PAW overlap for closed shells and is generalized as in Eq.~\eqref{eq:redistributedscalar} to remove open-shell redundancy.
The advantage of KPM is that it is numerically stable, being defined by a recursive relation 
with eigenvalues $< 1$, but it may be slower to reach a given resolution because it 
methodically adds information over the entire spectrum. In fact, in analogy with the Fourier Transform, 
the Chebyshev components span the whole spectral range, so that the number of iterations 
is proportional to the ratio between the spectral range and the required resolution, 
even when the final spectrum is sharply peaked in a narrow subrange. 
In contrast, the Lanczos tridiagonalization may converge rapidly when the spectral weight 
of the excited wavefunction is confined to a narrow region, but the numerical orthogonality 
of the generated Krylov space is more fragile. Both methods are implemented.

\paragraph*{Transition operator and $Q$ dependence.}
For nonresonant inelastic x-ray scattering (XRS/NIXS), the measured intensity is proportional to the dynamic structure factor,
\begin{equation}
  S(\mathbf Q,\omega)=\sum_f \big|\langle f|\widehat O(\mathbf Q)|g\rangle\big|^2\,
  \delta\!\big[\omega-(E_f-E_g)\big],
  \label{eq:SQomega}
\end{equation}
with $\widehat O(\mathbf Q)=\sum_j e^{i\mathbf Q\cdot \mathbf r_j}$ the Fourier component of the electron density (restricted in practice to the core-level excitation subspace on the absorber in our PAW embedding).
In nonresonant IXS/XRS the photon polarizations enter the double-differential cross section only through the usual Thomson (kinematic) prefactor (e.g.\ a factor $|{\boldsymbol\epsilon}_{\mathrm{in}}\!\cdot{\boldsymbol\epsilon}_{\mathrm{out}}|^2$ or its polarization-averaged form when the scattered polarization is not analyzed), whereas the many-body excitation operator is the density Fourier component $\widehat O(\mathbf Q)$. Since we focus on normalized spectral shapes, this prefactor is omitted here.
Using $\delta(x)=\frac{1}{2\pi}\int dt\, e^{ixt}$ and defining the excited wavepacket
$|\widehat\psi_0(\mathbf Q)\rangle=\widehat O(\mathbf Q)|g\rangle$ leads to Eq.~\eqref{eq:spectra}.
Therefore, all selection rules and the full $\mathbf Q$ dependence enter through the construction of
$|\widehat\psi_0(\mathbf Q)\rangle$, while the propagator itself is $\mathbf Q$-independent.
Expanding $e^{i\mathbf Q\cdot\mathbf r}$ into spherical Bessel functions and spherical harmonics makes explicit that low $Q$ emphasizes dipole-like ($L=1$) channels, while increasing $Q$ progressively enhances higher multipoles.

\paragraph*{Meaning of the symbols and numerical organization.}
In Eq.~\eqref{eq:initialwf} the ket on the \emph{left} of each tensor product, $|a_i\rangle$, is a \emph{Fock-space} basis vector for the absorber’s localized subspace. Concretely, it is specified in second quantization by the binary occupations ($0/1$) of all relevant on-site spin–orbitals $(n,\ell,m,\sigma)$—and, when present, of additional radial channels associated with the same $\ell$ (e.g., $4f$, $5f$, $6f$ or $6p$ arising from multiple PAW projectors). In this sense $|a_i\rangle$ compactly encodes an \emph{ionic configuration} (including the core hole) without any real-space expansion: it is a discrete label that fixes which localized spin–orbitals are occupied in that channel.%
\footnote{The spin of on-site (ionic) electrons is part of $|a_i\rangle$; the spin of the \emph{photoelectron} is carried by the channel index $i$ in the PW factor, as discussed below.}
The ket on the \emph{right}, $|\psi^i\rangle$, is instead a plane-wave (PW) pseudowavefunction living in the one-electron space of the photoelectron. Unlike the ionic factor, it is a freely evolving object: its components change in time under the entangled dynamics introduced below, and it can acquire any spatial form allowed by the PW representation. The index $i$ therefore labels \emph{channels} formed by an ionic basis state tensored with an associated PW wavefunction; by construction the sum in Eq.~\eqref{eq:initialwf} runs over all relevant ionic labels (and both spin channels).

From a computational standpoint, all PW wavefunctions $\{|\psi^i\rangle\}_{i=1}^{2M}$ are handled \emph{as a block}. We reuse the existing Quantum ESPRESSO infrastructure that applies linear operations to sets of PW states (as commonly done for multiple bands); here, the same batched machinery acts on \emph{channels} rather than on eigenbands. In practice, the “band index” of the PW engine is repurposed to enumerate the $2M$ entangled channels $|\psi^i\rangle$ attached to the different ionic labels $|a_i\rangle$. 

The coupling between channels—responsible for mixing amplitudes associated with different ionic configurations—is provided by precomputed scattering matrices imported from a multiplet code. These matrices encode the \emph{local atomic physics}, namely Coulomb multipole interactions and spin–orbit coupling.

\paragraph*{Local charge fluctuations in the open shell.}
The entangled form in Eq.~\eqref{eq:initialwf} also implies that the method naturally explores \emph{local charge fluctuations} on the absorber. Concretely, consider an ionic label whose open shell contains \(n\) electrons. Whenever the associated PW factor carries nonzero amplitude on the absorber’s open-shell orbitals, the \emph{local} occupancy of that shell is effectively \(n{+}1\): one electron comes from the ionic label and one from the PW component that momentarily resides in the localized channel. Conversely, when the PW amplitude has no overlap with those localized orbitals, the shell remains at occupancy \(n\). As the PW wavefunctions evolve, amplitude is continuously transferred between these two situations.

\paragraph*{Active orbitals of the Hilbert space.}
The Hilbert space is constructed by the multiplet code, together with the scattering factors.
It is determined by the user-choice concerning the core shell ( $4d$ in our examples), the open shell ( here $4f$) and its base occupancy $n$
(here $n=0,1$), and the complementing active orbitals which in our examples are $5f$, $6f$, and $6p$. They must correspond to existing PAW orbitals on whose basis the multiplet code computes matrix elements, using the  all-electrons wave-functions. This procedure is further illustrated by the worked out examples which come with the new version of XSPECTRA integrated in the QE suite.~\cite{xspectruplet}

In constructing the Hilbert basis, we impose the following constraints: the occupancy in the open shell fluctuates between $n$ and $n+1$; the core shell contains one hole;  the remaining active shells contain zero or one electrons; and finally  the total number of electrons fluctuates between $n$ and $n+1$, this without counting the core shell which has one hole.

\paragraph*{PW  plane-wave single-particle base model and spin treatment.}
In the present implementation, the plane-wave single-particle basis is obtained from a \emph{spinless} self-consistent (SCF) calculation. The underlying potential and the PAW projectors define a single-particle framework that does not carry an explicit spinor structure. The spin degree of freedom is instead introduced at the level of the entangled Hamiltonian, through the scattering terms imported from the multiplet code. These matrices include both Coulomb multipole interactions and spin–orbit coupling, so that the relevant spin physics—particularly the core and open-shell splittings that shape the experimental spectra—is incorporated via the atomic part of the problem. This design reflects the fact that, in core-level x-ray spectroscopies, the experimentally visible spin dynamics originates almost entirely from near-core, localized effects, whereas the spin behaviour of the extended photoelectron plays only a secondary role. 

In future developments, nothing prevents extending the framework to use a fully spinor plane-wave single-particle base model. A non-collinear (or fully relativistic) SCF calculation would naturally provide two-component spinor wavefunctions, allowing one to describe spin–orbit effects also in the delocalized region, while retaining the same entangled construction presented here.

\paragraph*{Why a Pauli projector is needed.}
In each entangled channel $|a_i\rangle\otimes|\psi^i\rangle$, the PW factor $|\psi^i\rangle$ is free to explore the full plane‑wave space. When we look at its \emph{local} content on the absorbing atom via PAW, nothing a priori prevents $|\psi^i\rangle$ from acquiring a finite projection onto an orbital that is \emph{already occupied} in the ionic labeled configuration $|a_i\rangle$. That would amount to placing two indistinguishable electrons (with the same spin as channel $i$) on the same localized orbital—i.e., a violation of Pauli. We therefore enforce, for each channel, a Pauli projector that removes from $|\psi^i\rangle$ any component parallel to the subspace spanned by the occupied local orbitals of $|a_i\rangle$.

\paragraph*{PAW notation.}
Within PAW we denote by $|\beta_\eta\rangle$ the projectors centered on the absorber (distinguished by angular momentum and, if needed, radial index), and by $|\phi_\eta\rangle$ the corresponding pseudo partial waves that are \emph{dual} to the projectors, i.e.\ $\langle \beta_\eta | \phi_{\eta'} \rangle = \delta_{\eta\eta'}$. These provide localized channels to test and manipulate the amplitude of $|\psi^i\rangle$ on specific atomic‑like orbitals (e.g., the $4f$ manifold). Because we use non‑norm‑conserving pseudopotentials, scalar products in PW space are taken with the overlap operator $S^{\mathrm{PW}}$.

\paragraph*{Definition and properties.}
Let $\mathrm{Occ}(a_i)$ be the set of PAW channels corresponding to orbitals that are occupied in the ionic state $|a_i\rangle$ \emph{with the same spin as channel $i$}. The Pauli projector acting on $|\psi^i\rangle$ is
\begin{equation}
  P^p_i \;=\; I \;-\; \sum_{\eta \in \mathrm{Occ}(a_i)}
  \frac{\,|\phi_\eta\rangle \langle \phi_\eta|\, S^{\mathrm{PW}}\,}{\langle \phi_\eta |\, S^{\mathrm{PW}}\, | \phi_\eta \rangle}\, .
  \label{eq:PauliProjector}
\end{equation}
In words, $P^p_i$ subtracts from $|\psi^i\rangle$ its component along each occupied local orbital, where “component” and “orthogonality” are meant in the $S^{\mathrm{PW}}$ metric appropriate for PAW. All other components are left unchanged. In the Cerium case considered here, the sum typically runs over the occupied $4f$ orbitals (and may be extended to any additional local shells explicitly included).

\paragraph*{Application during time evolution.}
We apply $P^p_i$ whenever the formal evolution operator is applied to the (entangled) excited wavepacket (e.g., within the Lanczos or KPM recursion). This guarantees that, even if the dynamics mixes channels, no PW factor $|\psi^i\rangle$ can ever develop a nonzero projection onto an orbital that is already occupied in its ionic label $|a_i\rangle$.
For the full entangled state, we use the block‑diagonal operator
\begin{equation}
  \widehat{P}^p \;=\; \sum_{i=1}^{2M} |a_i\rangle \langle a_i| \, P^p_i,
\end{equation}
which enforces Pauli independently in each channel.

%% \noindent
%% \emph{Remark.} The spin selectivity in $\mathrm{Occ}(a_i)$ is essential: Pauli forbids double occupation in the \emph{same spin channel}. Opposite‑spin occupation of the same spatial orbital is allowed and is handled consistently because the spin label is carried by the channel index $i$.

\paragraph*{Redundancy and scalar product.}
The entangled representation in Eq.~\eqref{eq:initialwf} is deliberately redundant whenever the photoelectron has nonzero amplitude on the \emph{open-shell} atomic subspace. Intuitively, a ionic configuration in which  $(n{+}1)$ electrons sit on an open shell  can be realized in $n{+}1$ equivalent ways by deciding which of the indistinguishable electrons we call “the photoelectron.” Our tensor-product construction materializes these $n{+}1$ choices as $n{+}1$ different channels $|a_i\rangle\!\otimes\!|\psi^i\rangle$. As a result, the \emph{atomic-projected part} of the state would be counted multiple times in naive norms and overlaps, unless we correct for this redundancy.

\paragraph*{A concrete example ($n=1$).}
Consider an ion with a single $4f$ electron (\(n=1\)). Let one ionic label be $|a_{(z^3)}\rangle$, meaning that the $4f$ electron occupies (schematically) the $z^3$ orbital. If  $|\phi_{xyz}\rangle$ is the PAW pseudo partial wave corresponding to a $4f$ orbital, the product $|a_{(z^3)}\rangle\!\otimes\!|\phi_{xyz}\rangle$ represents one way of forming a two-electron $4f$ configuration with orbitals $\{z^3, xyz\}$. But the \emph{same physical configuration} also appears in a different channel whose ionic label is $|a_{(xyz)}\rangle$ and whose PW factor carries the $z^3$ component. Thus, when we sum over channels, the atomic-projected contribution associated with $\{z^3, xyz\}$ is counted \emph{twice}. For a general open shell with occupancy $n$, any $(n{+}1)$-electron configuration obtained by placing the photoelectron in the open-shell subspace is represented \emph{$n{+}1$ times} across the entangled sum.

This is why only the part of $|\psi^i\rangle$ that lives in the open-shell atomic subspace needs a correction: components orthogonal to that subspace (i.e., truly delocalized or with different $l$) are unique and are not overcounted.

\paragraph*{Projector onto the open-shell subspace.}
To isolate the redundant piece we introduce the PAW projector
\begin{equation}
  P^a \;=\; \sum_{\eta\in\mathrm{o.s.}} 
  \frac{\,|\phi_\eta\rangle\langle\phi_\eta|\, S^{\mathrm{PW}}\,}
       {\,\langle \phi_\eta |\, S^{\mathrm{PW}}\, | \phi_\eta \rangle\,},
  \label{eq:PaProjector}
\end{equation}
where the sum runs over all PAW channels ($\eta$) that span the \emph{open-shell} on the absorber (e.g., the $4f$ manifold in Ce), $|\phi_\eta\rangle$ are the dual pseudo partial waves, and $S^{\mathrm{PW}}$ is the PW overlap operator appropriate for PAW. The operator $P^a$ extracts, from any PW state, precisely the localized component that gives rise to redundancy.

\paragraph*{Redundancy-corrected scalar product.}
With this notation, a redundancy-safe norm (and, more generally, overlap) for the entangled state is obtained by summing the standard PW metric over channels and then subtracting the overcounted atomic-projected weight with the correct combinatorial factor. Explicitly,
\paragraph*{Single open shell with fixed $n$.}
A redundancy-safe scalar product is then
\begin{equation}
  \langle \widehat\psi | \widehat S | \widehat\psi \rangle
  \;=\; \sum_{i=1}^{2M} \!\left(
    \langle \psi^i | S^{\mathrm{PW}} | \psi^i \rangle
    - \frac{n}{n{+}1}\, \langle \psi^i | P^{a} S^{\mathrm{PW}} P^{a} | \psi^i \rangle
  \right),
  \label{eq:redistributedscalar}
\end{equation}
The coefficient $n/(n{+}1)$ does exactly what the example suggests:
- for $n=0$ (closed shell, e.g.\ Ce$^{4+}$) there is no redundancy and the correction vanishes;
- for $n=1$ each atomic-projected piece is present twice, so we subtract one half of its total weight, leaving a single physical copy;
- in general, every $(n{+}1)$-electron open-shell configuration appears $n{+}1$ times, and subtracting $n/(n{+}1)$ of its total restores a single representation.

\paragraph*{PW Hamiltonian and Ionic Hamiltonians}
The one-particle PW Hamiltonian $H^{\mathrm{PW}}$ (from QE) governs the photoelectron dynamics; for non-norm-conserving PPs~\cite{vanderbilt} the generalized Schr\"odinger equation, without considering the ionic Hamiltonians but only the QE part, reads
\begin{equation}
  \partial_t^{QE} |\psi\rangle = -i \left(S^{\mathrm{PW}}\right)^{-1} H^{\mathrm{PW}} |\psi\rangle.
\end{equation}

which, together with Lanczos tridiagonalization to compute the spectra, is the basis of the former XSPECTRA implementation. 
Here we consider, beside this term,  also the action of the ionic Hamiltonian which we split  into  $H^{A_{n+1}}$, acting on the PAW projections of $|\psi^i\rangle$ over the open shell plus extra projectors,  and $H^{A_n}$ for the sector where the photoelectron does not project over the ionic orbitals. Their matrix elements are precomputed with a multiplet code (Hilbert++~\cite{vanadiohilb,amarcord})  considering  electron-electron Coulomb interaction and spin-orbit. %% ricordati il double counting
The relevant sources are distributed with QE.

The Hamiltonian $\widehat H^{A_{n+1}}$  is:
\begin{equation}
\widehat H^{A_{n+1}} =  \sum_{\kappa,i,j,\eta}
    \left |  \beta_\kappa \right >   \left| a_i \right\rangle
    \frac{h_{\kappa,i,j,\eta}}{f_\kappa f_\eta}
    \left\langle  a_j \right | \left\langle \beta_\eta \right |,
\end{equation}
where the sum over $\kappa,\eta$ spans all PAW projectors.
The factors $f_\kappa$ and $f_\eta$ compensate the redistribution induced by $\widehat R$ (defined in Eq.~\eqref{eq:R_operator}) in the open-shell sector:
\begin{equation*}
  f_\xi =
  \begin{cases}
    n+1, & \xi \in \mathrm{o.s.},\\
    1,   & \text{otherwise},
  \end{cases}
  \qquad (\xi=\kappa,\eta).
\end{equation*}
Consequently, terms involving two open-shell projectors carry an overall prefactor $1/(n+1)^2$,
while terms with exactly one open-shell index (i.e., $\kappa\in \mathrm{o.s.}$ and $\eta\notin\mathrm{o.s.}$ or vice versa) carry $1/(n+1)$.
In short the $h_{\kappa,i,j,\eta}$ coefficients describe the many-body interactions of the ion and the photoelectron considered together.

The Hamiltonian $\widehat H^{A_{n}}$ acts on the subspace where the photoelectron does not project on the absorbing site
\begin{equation}
  \widehat H^{A_{n}} =  \sum_{i,j}     \left| a_i \right\rangle h^\prime_{i,j} \left\langle  a_j \right | \,
  \Big( S^{\mathrm{PW}} -  \sum_{\eta} \left | \beta_\eta \right >\left\langle \beta_\eta \right | \Big),
\end{equation}

This operator satisfies two requirements: i) its expectation value is zero for a wavefunction whose PW part is a linear combination of
$|\phi_\eta\rangle$ (which are dual to projectors); ii) it recovers its {\it n-occupancy} limit when the PW wavefunction has no projection on the ion.
In short the $h^\prime_{i,j}$ coefficients describe the many-body interactions within the ion without the photoelectron.

\paragraph*{Redistribution operator.}
As discussed above, the atomic-projected part of a given $(n{+}1)$-electron open-shell configuration appears $n{+}1$ times across different channels. Our \emph{ansatz} is that all these \emph{equivalent representations} carry the \emph{same} complex amplitude at any time, i.e., the coefficient of a physical configuration must not depend on how we label “which electron is the photoelectron.” To enforce this indistinguishability in the entangled representation at each application of the time-evolution operator  we apply to its result the following operator:

\begin{equation}
  \widehat R = I + \sum_{a_i, \eta \in \mathrm{o.s.}} \;
  \sum_{ (a_j, \kappa) \equiv^\prime (a_i, \eta) }
     s_{j\kappa i\eta}
    \left| a_j \right\rangle
      \frac{\left |  \phi_\eta \right\rangle \left\langle \phi_\kappa \right | S^{\mathrm{PW}} \left\langle  a_i  \right| }
    { \big\| \phi_\eta \big\|^{1/2} \big\| \phi_\kappa \big\|^{1/2} } \,,
  \label{eq:R_operator}
\end{equation}
where the terms appearing in the denominator are defined as $\|\phi_\eta\| \equiv \langle \phi_\eta | S^{\mathrm{PW}} | \phi_\eta\rangle$, and the condition $(a_j,\kappa)\equiv^\prime  (a_i,\eta)$ means that the two channel–projector labels are two different but equivalent representations of the \emph{same} $(n{+}1)$-electron open-shell configuration. The sign $s_{j\kappa i\eta}$ enforces the correct Fermi statistics. The operator acts only in the atomic-projected sector spanned by the PAW partial waves $\{|\phi_\eta\rangle\}$ corresponding to the open shell ($\eta \in \mathrm{o.s.}$).

\noindent
\emph{Why it is needed (illustrative view).} Consider the $n{=}1$ case with two distinct $4f$ orbitals, say $z^3$ and $xyz$ (as in the example above). The same local two-electron configuration $\{z^3,xyz\}$ appears in two channels:
(i) the channel with ionic label $|a_{(z^3)}\rangle$ whose PW factor has a component along $|\phi_{xyz}\rangle$,
and (ii) the channel with ionic label $|a_{(xyz)}\rangle$ whose PW factor has a component along $|\phi_{z^3}\rangle$.
During time evolution, scattering may create (or modify) the amplitude of $\{z^3,xyz\}$ \emph{asymmetrically} in these two channels
%%—for instance, the update may enter through the first channel because its PW wavefunction happens at that moment to carry more local weight, while the second channel could represent a photoelectron far from the absorber.
If left unchecked, the two “copies” of the \emph{same} physical configuration would drift to different coefficients purely due to labelling. The action of $\widehat R$ restores equality of those coefficients by redistributing the increment from one copy to all its equivalents, so that the configuration $\{z^3,xyz\}$ is represented with a single, consistent amplitude, independent of channel labels.

\noindent
\emph{Operational use.} We apply $\widehat R$ after each application of the formal evolution operator to the excited wavepacket $\left|\widehat{\psi}_0\right\rangle$ (e.g., within the Lanczos or KPM recursion): we act on the block of PW channels, enforce Pauli in each channel, then symmetrize with $\widehat R$. This keeps the redundancy-corrected scalar product consistent and guarantees that numerical results do not depend on arbitrary choices in channel labelling.
%% In the fully developed formalism, the local scattering terms are compensated for this symmetrization in advance by appropriate normalization factors (introduced later), so that matrix elements are neither double-counted nor diluted by the averaging.
To compensate for the demultiplexing effect of the redistribution operator on the strength of matrix elements, we use the factors $f_\kappa$ and $f_\eta$ in $\widehat H^{A_{n+1}}$ as defined above.

\paragraph*{Entangled time-evolution.}
Based on the above defined operators,
the time evolution of the entangled state is
\begin{align}
  \partial_t \left | \widehat \psi \right > &= \widehat D_t \left | \widehat \psi \right >, \\
  \widehat D_t &= -i\, \widehat R \,\widehat P^{p}\, \left(\widehat S^{\mathrm{PW}}\right)^{-1}
  \left(\widehat H^{\mathrm{PW}} + \widehat H^{A_{n}} + \widehat H^{A_{n+1}}\right).
  \label{eq:beyond}
\end{align}
Here    $\widehat S^{\mathrm{PW}}$ and $\widehat H^{\mathrm{PW}}$ are direct sum, one of $S^{\mathrm{PW}}$ and the other of $H^{\mathrm{PW}}$, over the ionic channels.
This equation, together with Eq.~\ref{eq:spectra} defines the spectra.

%% % ============================================================
%% \section{Implementation (XSPECTRUPLET)}
%% % ============================================================

%% We implemented the above formalism by extending the XSPECTRA module~\cite{bunau,mauri1,mauri2} of Quantum ESPRESSO~\cite{QE-2017,QE-2009}

%% into a new code, \emph{XSPECTRUPLET}. The code:
%% (i) builds replicated PW channels $|\psi^i\rangle$ tied to ionic basis states $|a_i\rangle$;
%% (ii) enforces Pauli exclusion via $P_i^p$ at every propagation step;
%% (iii) applies ionic scatterings through $H^{A_{n}}$ and $H^{A_{n+1}}$ constructed from Hilbert++~\cite{vanadiohilb,amarcord};
%% (iv) advances in time with the operator $\widehat D_t$ in Eq.~\eqref{eq:beyond}, using either KPM~\cite{kernel} or Lanczos~\cite{dagotto} to obtain $f(\omega)$ in Eq.~\eqref{eq:spectra}.

%% Coupling to PAW is crucial: PAW projectors provide localized channels to (a) project PWs onto atomic subspaces (open shell and extra projectors), (b) embed precomputed ionic matrix elements $h_{\kappa i j \eta}$ and $h_{ij}$, and (c) guarantee consistency with the generalized PW scalar product $S^{\mathrm{PW}}$.

%% % -- Quick-start pointer (full user guide goes to SM) --
%% % For a concise Quick-start summary and full runnable examples (inputs + commands),
%% % see Supplemental Material, Secs. S2–S5.  %% TODO: add SM with inputs and scripts.

% ============================================================
\section{Computational setup}
% ============================================================

The calculation begins with the evaluation of the Slater integrals involving the core-hole orbital, the open-shell wave functions, and the all-electron wave functions associated with those pseudopotential projectors that are included in the definition of the ionic Hamiltonians. When the core-hole orbital is not part of the pseudopotential, it is generated by a self-consistent spherical-atom calculation, which also provides the corresponding spin–orbit coupling strength.

With these ingredients, a straightforward multiplet calculation is performed to determine the ionic ground state. The initial (entangled) excited wavepacket  is then constructed by applying the excitation operator to this ground state and subsequently projecting the resulting state onto the plane-wave representation through the PAW wave functions.

To provide an accurate description of the crystal environment, the ionic Hamiltonians also require the crystal-field contribution. These ingredients—the crystal field and the corresponding density matrix—can optionally be extracted from a preliminary SCF calculation followed by a Wannierization of the open-shell orbitals, together with the rotation of both the one-particle Hamiltonian and the density matrix into the Wannier basis. Using the resulting open-shell density matrix, we then remove the double-counting term associated with the non-spherical mean-field electron–electron interaction within the open shell. This correction is necessary to avoid counting this contribution twice: once through the crystal-field term and again through the explicit many-body interactions.

Alternatively, in the example discussed below, we adopt empirical tight-binding Slater–Koster parameters~\cite{vanadiohilb} to model the crystal field in the ground-state multiplet calculation. The resulting density matrix is then used to subtract the double-counting contribution from the \( h_{\kappa,i,j,\eta} \) and \( h'_{i,j} \) coefficients before performing the \textit{xspectruplet} calculation.

The crystal field plays an important role not only through its overall strength but, more importantly, through its ability to orient the open-shell orbitals in the ground state. When these orbitals are partially occupied, such orientation may induce an angular dependence in the spectra. In our test case, which concerns the highly localized \(4f\) shell, we employed an arbitrary crystal-field strength of less than one tenth of an eV in the Slater–Koster parametrization. This magnitude is sufficient to enforce a well-defined orbital orientation once the crystallographic positions of the neighbouring oxygen atoms are provided to the multiplet code.

The spherical component of the ionic Hamiltonians is then made neutral with respect to the plane-wave Hamiltonian. This is achieved by setting the trace of \( h'_{i,j} \) to zero, while for the coefficients \( h_{\eta,i,j,\eta} \) the offset is applied separately for each value of \( \eta \).

\paragraph*{Convergence and resolution control (Lanczos vs KPM).}
In practice, evaluating Eq.~\eqref{eq:spectra} requires truncating the iterative expansion.
In the Lanczos approach, the correlation function is represented by a continued fraction obtained from the Krylov subspace generated by $\widehat D_t$ acting on $\left|\widehat{\psi}_0\right\rangle$; the recursion is therefore stopped after $N_{\mathrm{L}}$ steps.
Convergence is assessed empirically by monitoring the change of the spectrum upon increasing $N_{\mathrm{L}}$ (e.g., requiring that $f_{N_{\mathrm{L}}+\Delta N}(\omega)$ differs from $f_{N_{\mathrm{L}}}(\omega)$ by less than a user-defined threshold over the energy window of interest).
As is well known, Lanczos can suffer from a progressive loss of orthogonality in finite precision arithmetic, which may lead to numerical instabilities or spurious features if pushed to very large $N_{\mathrm{L}}$. This can be mitigated by full or selective reorthogonalization, at the price of additional computational and  memory overhead, which can become prohibitive in large-scale HPC runs.

In the kernel polynomial method (KPM), the effective operator is first rescaled so that its spectrum lies within the domain of the Chebyshev recursion, and the spectral resolution is then controlled \emph{deterministically} by the expansion order $N_{\mathrm{C}}$ (together with the chosen kernel/broadening).
Thus, for a target resolution $\Delta E$ over a fixed spectral range, $N_{\mathrm{C}}$ is fixed a priori, and the procedure does not rely on an empirical stopping criterion.
In practice, the two methods can be used in a complementary way: one may (i) use KPM alone when a guaranteed target resolution is desired, or (ii) use KPM as a numerically stable reference to calibrate the minimum $N_{\mathrm{L}}$ needed by Lanczos to reach the same effective resolution and lineshape, thereby providing an a posteriori validation of Lanczos stability.

% ============================================================
\section{Results and validation}
% ============================================================

\begin{figure}[t]
  \centering
  \includegraphics[width=\linewidth]{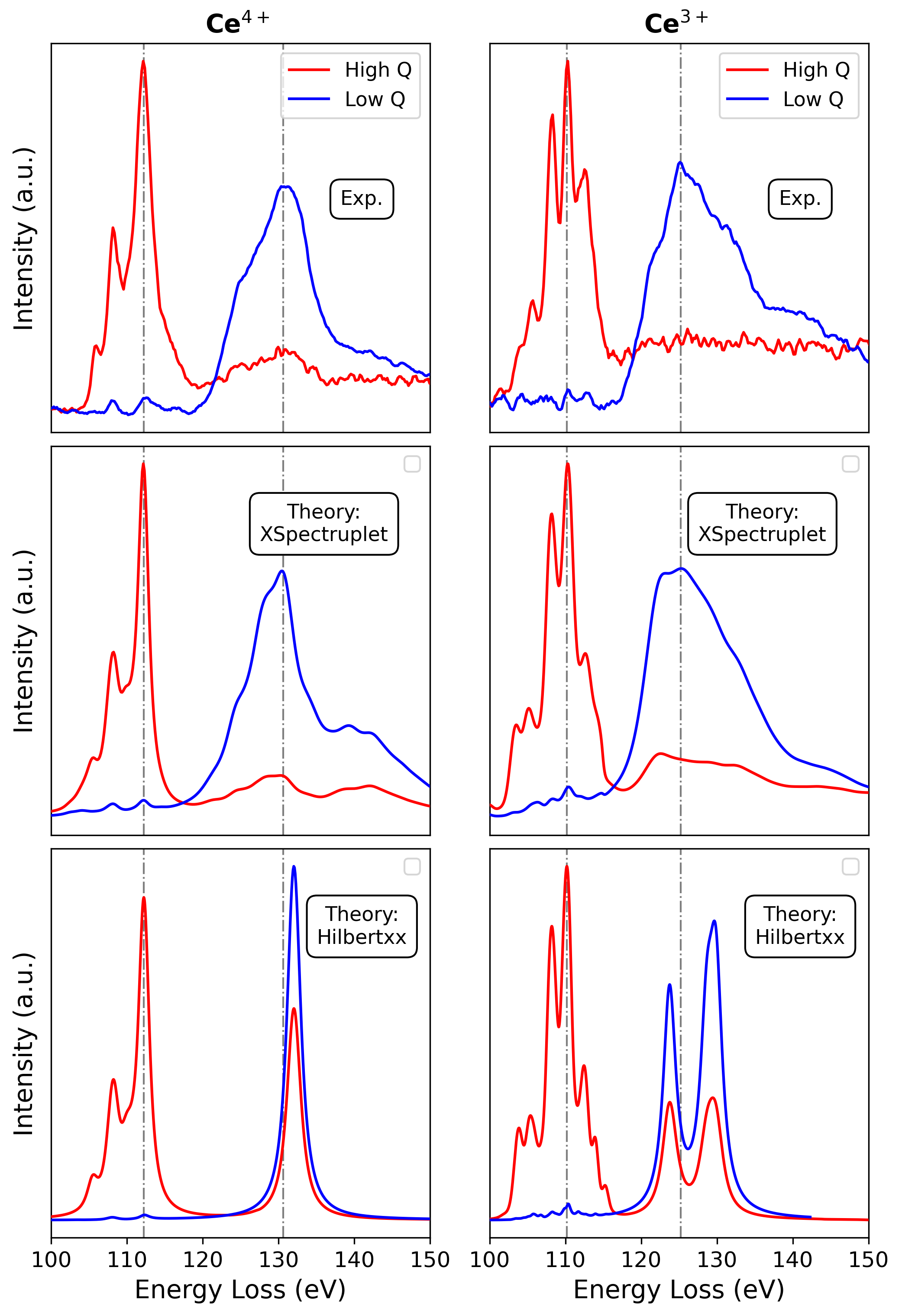}
  \caption{XRS at the Ce $N_{4,5}$ edge for CeO$_2$ (left) and $\mathrm{Ce_2(SO_4)_3}$~\cite{Das2024} (right): experiment (top), xspectruplet (middle), multiplet (bottom). Red: high-$Q$ ($9.5$~\AA$^{-1}$); blue: low-$Q$ ($3.5$~\AA$^{-1}$). The entangled propagation reproduces localized peaks and continuum GDR structure with correct energy and lineshape, including low-energy shoulders otherwise absent in pure multiplet.}
  \label{fig:plot1}
\end{figure}

We show in Fig.~\ref{fig:plot1} the experimental (top row) and computed (middle row)  XRS spectra at the N$_{4,5}$ edge of cerium for $\mathrm{CeO_2}$ and cerium(III) sulfate ($\mathrm{Ce_2(SO_4)_3}$)~\cite{Das2024}. In the bottom line of Fig.~\ref{fig:plot1} the spectra have been calculated switching off the PW Hamiltonian, and considering only the $4f$ projectors and pseudo-functions, thus reproducing a simple multiplet calculation.

\paragraph*{Independent-particle baseline and why it is not shown.}
A natural reference would be a one-particle plane-wave/PAW spectrum as provided by the standard XSPECTRA workflow. 
However, in its current form XSPECTRA is not directly applicable to the Ce \(N_{4,5}\) thresholds with the PAW datasets employed here, and extending it to this edge would require additional dedicated setup for the \(4d\) core channel. 
More importantly, even if such an independent-particle calculation were performed, the resulting lineshape would largely reflect (i) the single-particle crystal-field splitting of the \(4f\)-derived final states (typically sub-eV) and (ii) the \(4d\) spin--orbit separation between the \(N_5\) and \(N_4\) components (of order a few eV), while missing the correlated \(4d^{-1}4f^{n+1}\) multiplet manifold and the resonance--continuum mixing that redistribute oscillator strength over a \(\gtrsim 20\)~eV window and produce the giant dipole resonance. 
For this reason, we use as controlled baselines the purely local multiplet limit obtained by switching off \(\widehat H^{\mathrm{PW}}\) (bottom panels of Fig.~\ref{fig:plot1}) and the full entangled propagation including \(\widehat H^{\mathrm{PW}}\) (middle panels).

The first compound contains $\mathrm{Ce~4f^0}$ ions, for which Eq.~\ref{eq:beyond} propagates two particles: the core hole and the photoelectron. 
The second compound instead contains $\mathrm{Ce~4f^1}$ ions, for which Eq.~\ref{eq:beyond} describes an entangled three-body final state (core hole, open-shell electron, and photoelectron), i.e., a multichannel extension of the standard core-exciton treatment.

Given the close chemical similarity and the nearly identical ionic radii of trivalent Ce and Pr, we used the experimentally established structure of $\mathrm{Pr_2(SO_4)_3}$ as a proxy for $\mathrm{Ce_2(SO_4)_3}$.
The graph shows the energy loss spectra for an exchanged $Q$ of $9.5\AA^{-1}$ (red line, high $Q$) and $3.5\AA^{-1}$ (blue line, low $Q$). 
The procedure to reproduce the calculation with the Quantum-ESPRESSO package is provided in the examples for the  {\it xspectruplet} mode which accompany the new QE sources~\cite{xspectruplet}.  
In all the calculations the Slater integrals are rescaled by a reduction factor equal to $0.85$ for all the Slater integrals in the case of Ce$^{4+}$, and equal to $0.75$ for Ce$^{3+}$.
The calculations are convoluted with a FWHM=$1.5$ eV Lorentzian for the high-Q part and with FWHM=$3$ eV  for the low-Q part. This choice was motivated by the agreement with the experimental shape.

Concerning Ce$^{4+}$, the extra shoulder in the low energy part of the resonance, emerges naturally as a result of the formalism.
This is an important result, which clarifies how to interpret correctly these features in relation to the oxidation state of the system.
% Without this correct interpretation the shoulders at lower energy could be taken  as the evidence of a chemically shifted peak of a reduced ion.  %% gg 24

The high-Q part is well reproduced both in shape and in the energy positions of the peaks. These positions strictly determine the Slater reduction factor. With the above so determined Slater reduction factors ( one for Ce$^{4+}$ and another one for Ce$^{3+}$) the energy position of the GDR peak is well reproduced in the {\it xspectruplet} calculation but is found at higher energy in the simple multiplet calculation.
This is an effect of the GDR weight on components belonging to the continuum, all  having a minor overlap with $4d$ than  $4f$ has, hence a lesser exchange with $4d$. Similar effects, where the renormalization of the effective Slater integrals depends on the resonance, have been already observed in the literature
and ascribed to the hybridization with states in the continuum, reproducing the same behaviour  with a multiplet calculation where the atomic orbitals of the model Hamiltonian have been complemented by a discrete number of fictitious states\cite{sengupta}.  

% ============================================================
\section{Discussion and outlook}
% ============================================================

We have developed a quantum-mechanical formalism that treats the coherent electronic entanglement between a plane-wave photoelectron and a correlated manifold of localized ionic configurations of the absorber in \emph{ab initio} core-level spectroscopy calculations.

% The application of our QE-based implementation has allowed the correct interpretation of the GDR features in cerium oxide systems, in particular their shape and energy position.
% guadagno 4
Our method has allowed the correct interpretation of the GDR features in cerium oxide systems, in particular their shape and energy position,
% Our theoretical analysis and numerical implementation has the potential to shed new light on spectral regions that have so far been scarcely explored, opening up new avenues of investigation for systems whose spectral features lie in a “gray zone” between transitions to well-localized states and transitions to the continuum. In this frontier regime, the wavefunctions of the excited states can exhibit a hybrid nature: partially localized (as in discrete states) and partially extended (as in continuum channels).
% guadagno 5
and  has the potential to shed new light on 
% opening up new avenues of investigation for systems whose spectral features lie in a
% guadagno 11
the 
“gray zone” between transitions to well-localized states and transitions to the continuum.

% guadagno 28
% In this frontier regime, the wavefunctions of the excited states can exhibit a hybrid nature: partially localized (as in discrete states) and partially extended (as in continuum channels).

As shown in this work a particularly relevant example is the giant dipole resonance (GDR) in rare-earth elements, where electronic transitions involve deep-lying orbitals (often 4f and 5d), which show an intriguing interplay between tightly bound excitations and partial escape into the continuum. Similar phenomena also appear in uranium compounds
% where the high density of states associated with the 5f orbitals—together with their possible hybridization with ligand orbitals further enriches the landscape of electronic transitions.
% guadagno 14
%: part of the spectrum remains localized, while other excitations open up into continuum channels \cite{Ucompounds}.
and in  2p $\rightarrow$ 3d transition where  threshold effects and state mixing can make it challenging to distinguish between discrete energy levels and those that “spill over” into higher-energy continua.

% guadatno 36
%% This coexistence of discrete and continuum states—also typical of various autoionization processes, where resonant states compete with ionization pathways underscores the complexity of electron-hole interactions in a spectral regime that is still relatively unexplored.

%% guadagno 42
All these observations highlight how our approach, focused on exploring phenomena in this “hybrid” energy range, can pave the way for both the development of more refined theoretical models and the design of experiments aimed at exploiting these features rich spectroscopy domains.

% ============================================================
\section*{Data and code availability}
% ============================================================

%% TODO: Replace with the actual repository/DOI before submission.
All input files, scripts to reproduce the figures, and the {\it xspectruplet} source used in this work  are openly available as part of Quantum-ESPRESSO~\cite{xspectruplet}.

% ============================================================
\begin{acknowledgments}
% Use the standard APS acknowledgments environment.
  We thank Matteo Calandra, Joshua Elliot and Francesco Mauri for the advice and help during the integration of {\it xspectruplet} within  Quantum ESPRESSO, and the Quantum ESPRESSO community for providing a well-written and well-documented suite that enables smooth implementation of new ideas.
  %%%% the following text cannot be altered %%%%
  We acknowledge the European Synchrotron Radiation Facility (ESRF) for provision of synchrotron radiation facilities under proposal numbers IH-CH-1633 and IH-CH-1634 and we would like to thank F.~Gerbon for assistance and support in using beamline ID20. %
  %%% the preceding text cannot be altered %%%%
\end{acknowledgments}

% ============================================================
\bibliographystyle{apsrev4-2}
\bibliography{new_resonance}

@article{gordon,
doi = {10.1209/0295-5075/81/26004},
url = {https://dx.doi.org/10.1209/0295-5075/81/26004},
year = {2007},
month = {dec},
publisher = {},
volume = {81},
number = {2},
pages = {26004},
author = {R. A. Gordon and G. T. Seidler and T. T. Fister and M. W. Haverkort and G. A. Sawatzky and A. Tanaka and T. K. Sham},
title = {High multipole transitions in NIXS: Valence and hybridization in 4f systems},
journal = {Europhysics Letters}
}

@misc{xspectruplet,
  title        = {XSpectruplet Project},
  howpublished = {\url{https://gitlab.com/xspectruplet/xspectruplet}},
  note         = {Accessed: 2025-12-07},
  year         = {2025}
}

@article{quanty,
  title = {Multiplet ligand-field theory using Wannier orbitals},
  author = {Haverkort, M. W. and Zwierzycki, M. and Andersen, O. K.},
  journal = {Phys. Rev. B},
  volume = {85},
  issue = {16},
  pages = {165113},
  numpages = {20},
  year = {2012},
  month = {Apr},
  publisher = {American Physical Society},
  doi = {10.1103/PhysRevB.85.165113},
  url = {https://link.aps.org/doi/10.1103/PhysRevB.85.165113}
}

@article{bunau,
  title = {Projector augmented wave calculation of x-ray absorption spectra at the ${L}_{2,3}$ edges},
  author = {Bun\ifmmode \u{a}\else \u{a}\fi{}u, Oana and Calandra, Matteo},
  journal = {Phys. Rev. B},
  volume = {87},
  issue = {20},
  pages = {205105},
  numpages = {12},
  year = {2013},
  month = {May},
  publisher = {American Physical Society},
  doi = {10.1103/PhysRevB.87.205105},
  url = {https://link.aps.org/doi/10.1103/PhysRevB.87.205105}
}

@article{mauri2,
  title = {First-principles calculations of x-ray absorption in a scheme based on ultrasoft pseudopotentials: From $\ensuremath{\alpha}$-quartz to high-${T}_{c}$ compounds},
  author = {Gougoussis, Christos and Calandra, Matteo and Seitsonen, Ari P. and Mauri, Francesco},
  journal = {Phys. Rev. B},
  volume = {80},
  issue = {7},
  pages = {075102},
  numpages = {8},
  year = {2009},
  month = {Aug},
  publisher = {American Physical Society},
  doi = {10.1103/PhysRevB.80.075102},
  url = {https://link.aps.org/doi/10.1103/PhysRevB.80.075102}
}

@article{mauri1,
  title = {X-ray absorption near-edge structure calculations with the pseudopotentials: Application to the K edge in diamond and $\ensuremath{\alpha}$-quartz},
  author = {Taillefumier, Mathieu and Cabaret, Delphine and Flank, Anne-Marie and Mauri, Francesco},
  journal = {Phys. Rev. B},
  volume = {66},
  issue = {19},
  pages = {195107},
  numpages = {8},
  year = {2002},
  month = {Nov},
  publisher = {American Physical Society},
  doi = {10.1103/PhysRevB.66.195107},
  url = {https://link.aps.org/doi/10.1103/PhysRevB.66.195107}
}

@article{QE-2017,
  author={P Giannozzi and O Andreussi and T Brumme and O Bunau and M Buongiorno Nardelli
  and M Calandra and R Car and C Cavazzoni and D Ceresoli and M Cococcioni and N Colonna
  and I Carnimeo and A Dal Corso and S de Gironcoli and P Delugas and R A DiStasio Jr and A Ferretti
  and A Floris and G Fratesi and G Fugallo and R Gebauer and U Gerstmann and F Giustino and T Gorni
  and J Jia and M Kawamura and H-Y Ko and A Kokalj and E Küçükbenli and M Lazzeri and M Marsili
  and N Marzari and F Mauri and N L Nguyen and H-V Nguyen and A Otero-de-la-Roza and L Paulatto
  and S Poncé and D Rocca and R Sabatini and B Santra and M Schlipf and A P Seitsonen
  and A Smogunov and I Timrov and T Thonhauser and P Umari and N Vast and X Wu and S Baroni},
  title={Advanced capabilities for materials modelling with QUANTUM ESPRESSO},
  journal={Journal of Physics: Condensed Matter},
  volume={29},
  number={46},
  pages={465901},
  url={http://stacks.iop.org/0953-8984/29/i=46/a=465901},
  year={2017},
}

@article{QE-2009,
	Author = {Paolo Giannozzi and Stefano Baroni and Nicola Bonini and Matteo Calandra and Roberto Car
 and Carlo Cavazzoni and Davide Ceresoli and Guido L Chiarotti and Matteo Cococcioni and Ismaila Dabo
 and Andrea {Dal Corso} and Stefano de Gironcoli and Stefano Fabris and Guido Fratesi and Ralph Gebauer
 and Uwe Gerstmann and Christos Gougoussis and Anton Kokalj and Michele Lazzeri and Layla Martin-Samos
 and Nicola Marzari and Francesco Mauri and Riccardo Mazzarello and Stefano Paolini and Alfredo Pasquarello
 and Lorenzo Paulatto and Carlo Sbraccia and Sandro Scandolo and Gabriele Sclauzero and Ari P Seitsonen
 and Alexander Smogunov and Paolo Umari and Renata M Wentzcovitch},
	Journal = {Journal of Physics: Condensed Matter},
	Number = {39},
	Pages = {395502 (19pp)},
	Title = {QUANTUM ESPRESSO: a modular and open-source software project for quantum simulations of materials},
	Url = {http://www.quantum-espresso.org},
	Volume = {21},
	Year = {2009}}

@article{kernel,
  title={The kernel polynomial method},
  author={Wei{\ss}e, Alexander and Wellein, Gerhard and Alvermann, Andreas and Fehske, Holger},
  journal={Reviews of Modern Physics},
  volume={78},
  number={1},
  pages={275--306},
  year={2006},
  publisher={American Physical Society},
  doi={10.1103/RevModPhys.78.275}
}

@book{dagotto,
  author       = {Dagotto, Elbio},
  title        = {Correlated Electrons in High-Temperature Superconductors},
  publisher    = {Springer},
  year         = {1994},
  address      = {Berlin},
  isbn         = {978-3-642-08555-6},
  doi          = {10.1007/978-3-642-08555-6}
}

@article{amarcord,
  title = {Ligand-field atomic-multiplet calculations for arbitrary symmetry},
  volume = {61},
  ISSN = {1095-3795},
  url = {http://dx.doi.org/10.1103/PhysRevB.61.13540},
  DOI = {10.1103/physrevb.61.13540},
  number = {20},
  journal = {Physical Review B},
  publisher = {American Physical Society (APS)},
  author = {Mirone,  Alessandro and Sacchi,  Maurizio and Gota,  Susana},
  year = {2000},
  month = may,
  pages = {13540–13544}
}

@article{vanadiohilb,
author = {Longo, Alessandro and Wernert, Romain and Iadecola, Antonella and Sahle, Christoph J. and Stievano, Lorenzo and Croguennec, Laurence and Carlier, Dany and Mirone, Alessandro},
title = {An Original Empirical Method for Simulating V L2,3 Edges: The Example of KVPO4F and KVOPO4 Cathode Materials},
journal = {The Journal of Physical Chemistry C},
volume = {126},
number = {46},
pages = {19782-19791},
year = {2022},
doi = {10.1021/acs.jpcc.2c05334},

URL = { 
    
        https://doi.org/10.1021/acs.jpcc.2c05334
    
    

},
eprint = { 
    
        https://doi.org/10.1021/acs.jpcc.2c05334
    
    

}

}

@article{Das2024,
  title = {Deciphering the Ce3+ to Ce4+ Evolution: Insight from X‐ray Raman Scattering Spectroscopy at Ce N4, 5 Edges},
  ISSN = {1439-7641},
  url = {http://dx.doi.org/10.1002/cphc.202400742},
  DOI = {10.1002/cphc.202400742},
  journal = {ChemPhysChem},
  publisher = {Wiley},
  author = {Das,  Soumya K. and Longo,  Alessandro and Bianchi,  Eugenio and Bordenca,  Claudio V. and Sahle,  Christoph J. and Casaletto,  Maria Pia and Mirone,  Alessandro and Giannici,  Francesco},
  year = {2024},
  month = oct 
}

@article{Longo2023,
  title     = {Oxygen vacancy clusters in bulk cerium oxide and the impact of gold atoms},
  author    = {Longo, Alessandro and Mirone, Alessandro and De Clermont Gallerande, Emmanuelle and Sahle, Christoph J. and Casaletto, Maria Pia and Amidani, Lucia and Theofanidis, Stavros A. and Giannici, Francesco},
  year      = {2023},
  journal   = {Cell Reports Physical Science},
  volume    = {4},
  number    = {12},
  pages     = {101699},
  doi       = {10.1016/j.xcrp.2023.101699},
  publisher = {Elsevier},
  url       = {https://doi.org/10.1016/j.xcrp.2023.101699}
}

@article{kohn,
  title = {Self-Consistent Equations Including Exchange and Correlation Effects},
  author = {Kohn, W. and Sham, L. J.},
  journal = {Phys. Rev.},
  volume = {140},
  issue = {4A},
  pages = {A1133--A1138},
  numpages = {0},
  year = {1965},
  month = {Nov},
  publisher = {American Physical Society},
  doi = {10.1103/PhysRev.140.A1133},
  url = {https://link.aps.org/doi/10.1103/PhysRev.140.A1133}
}

@article{pbe,
  title = {Generalized Gradient Approximation Made Simple},
  author = {Perdew, John P. and Burke, Kieron and Ernzerhof, Matthias},
  journal = {Phys. Rev. Lett.},
  volume = {77},
  issue = {18},
  pages = {3865--3868},
  numpages = {0},
  year = {1996},
  month = {Oct},
  publisher = {American Physical Society},
  doi = {10.1103/PhysRevLett.77.3865},
  url = {https://link.aps.org/doi/10.1103/PhysRevLett.77.3865}
}

@article{Vinson2011,
  author    = {J. Vinson and J. J. Rehr and J. J. Kas and E. L. Shirley},
  title     = {Bethe-Salpeter Equation Calculations of Core Excitation Spectra},
  journal   = {Phys. Rev. B},
  volume    = {83},
  issue     = {11},
  pages     = {115106},
  year      = {2011},
  doi       = {10.1103/PhysRevB.83.115106},
  url       = {https://doi.org/10.1103/PhysRevB.83.115106}
}

@article{Prendergast2006,
  author    = {D. Prendergast and G. Galli},
  title     = {X-ray absorption spectra of water from first principles calculations},
  journal   = {Phys. Rev. Lett.},
  volume    = {96},
  pages     = {215502},
  year      = {2006},
  doi       = {10.1103/PhysRevLett.96.215502},
  url       = {https://doi.org/10.1103/PhysRevLett.96.215502}
}

@article{Vorwerk2017,
  author    = {C. Vorwerk and C. Cocchi and C. Draxl},
  title     = {Addressing electron-hole correlation in core excitations of solids: An all-electron many-body approach},
  journal   = {Phys. Rev. B},
  volume    = {95},
  pages     = {155121},
  year      = {2017},
  doi       = {10.1103/PhysRevB.95.155121},
  url       = {https://doi.org/10.1103/PhysRevB.95.155121}
}

@article{chemreview,
author = {Cohen, Aron J. and Mori-Sánchez, Paula and Yang, Weitao},
title = {Challenges for Density Functional Theory},
journal = {Chemical Reviews},
volume = {112},
number = {1},
pages = {289-320},
year = {2012},
doi = {10.1021/cr200107z},
note ={PMID: 22191548},
URL = { 
https://doi.org/10.1021/cr200107z
},
eprint = { 
  https://doi.org/10.1021/cr200107z
}
}

@article{gw,
  title={The GW method},
  author={Aryasetiawan, Ferdi and Gunnarsson, Olle},
  journal={Reports on progress in Physics},
  volume={61},
  number={3},
  pages={237},
  year={1998},
  publisher={IOP Publishing}
}

@article{dmft,
  title = {Dynamical mean-field theory of strongly correlated fermion systems and the limit of infinite dimensions},
  author = {Georges, Antoine and Kotliar, Gabriel and Krauth, Werner and Rozenberg, Marcelo J.},
  journal = {Rev. Mod. Phys.},
  volume = {68},
  issue = {1},
  pages = {13--125},
  numpages = {0},
  year = {1996},
  month = {Jan},
  publisher = {American Physical Society},
  doi = {10.1103/RevModPhys.68.13},
  url = {https://link.aps.org/doi/10.1103/RevModPhys.68.13}
}

@article{Yutsis,
  title={Mathematical apparatus of the theory of angular momentum},
  author={Yutsis, Adolfas P and Levinson, Ioshua Benʹi︠a︡minovich and Vanagas, Vladislavas Vladovich},
  journal={Academy of Sciences of the Lithuanian SS R},
  year={1962}
}

@misc{deGroot2008,
  title = {Core Level Spectroscopy of Solids},
  ISBN = {9780429195792},
  url = {http://dx.doi.org/10.1201/9781420008425},
  DOI = {10.1201/9781420008425},
  publisher = {CRC Press},
  author = {de Groot,  Frank and Kotani,  Akio},
  year = {2008},
  month = mar 
}

@article{w90,
title = {An updated version of wannier90: A tool for obtaining maximally-localised Wannier functions},
journal = {Computer Physics Communications},
volume = {185},
number = {8},
pages = {2309-2310},
year = {2014},
issn = {0010-4655},
doi = {https://doi.org/10.1016/j.cpc.2014.05.003},
url = {https://www.sciencedirect.com/science/article/pii/S001046551400157X},
author = {Arash A. Mostofi and Jonathan R. Yates and Giovanni Pizzi and Young-Su Lee and Ivo Souza and David Vanderbilt and Nicola Marzari}
}

@article{tt,
  title = {3d x-ray-absorption lines and the 3${d}^{9}$4${f}^{n+1}$ multiplets of the lanthanides},
  author = {Thole, B. T. and van der Laan, G. and Fuggle, J. C. and Sawatzky, G. A. and Karnatak, R. C. and Esteva, J.-M.},
  journal = {Phys. Rev. B},
  volume = {32},
  issue = {8},
  pages = {5107--5118},
  numpages = {0},
  year = {1985},
  month = {Oct},
  publisher = {American Physical Society},
  doi = {10.1103/PhysRevB.32.5107},
  url = {https://link.aps.org/doi/10.1103/PhysRevB.32.5107}
}

@article{vanderbilt,
  title = {Soft self-consistent pseudopotentials in a generalized eigenvalue formalism},
  author = {Vanderbilt, David},
  journal = {Phys. Rev. B},
  volume = {41},
  issue = {11},
  pages = {7892--7895},
  numpages = {0},
  year = {1990},
  month = {Apr},
  publisher = {American Physical Society},
  doi = {10.1103/PhysRevB.41.7892},
  url = {https://link.aps.org/doi/10.1103/PhysRevB.41.7892}
}

@article{keith,
title = {Efficient implementation of core-excitation Bethe–Salpeter equation calculations},
journal = {Computer Physics Communications},
volume = {197},
pages = {109-117},
year = {2015},
issn = {0010-4655},
doi = {https://doi.org/10.1016/j.cpc.2015.08.014},
url = {https://www.sciencedirect.com/science/article/pii/S0010465515003008},
author = {K. Gilmore and John Vinson and E.L. Shirley and D. Prendergast and C.D. Pemmaraju and J.J. Kas and F.D. Vila and J.J. Rehr}
}

@article{sengupta,
  title = {Coexistence of bound and virtual-bound states in shallow-core to valence x-ray spectroscopies},
  author = {Sen Gupta, Subhra and Bradley, J. A. and Haverkort, M. W. and Seidler, G. T. and Tanaka, A. and Sawatzky, G. A.},
  journal = {Phys. Rev. B},
  volume = {84},
  issue = {7},
  pages = {075134},
  numpages = {9},
  year = {2011},
  month = {Aug},
  publisher = {American Physical Society},
  doi = {10.1103/PhysRevB.84.075134},
  url = {https://link.aps.org/doi/10.1103/PhysRevB.84.075134}
}

@article{Onida2002RMP,
  author  = {Onida, G. and Reining, L. and Rubio, A.},
  title   = {Electronic excitations: density-functional versus many-body Green's-function approaches},
  journal = {Rev. Mod. Phys.},
  volume  = {74},
  pages   = {601},
  year    = {2002},
  doi     = {10.1103/RevModPhys.74.601}
}

@article{CudazzoReining2020PRR,
  author  = {Cudazzo, Pierluigi and Reining, Lucia},
  title   = {Correlation satellites in optical and loss spectra},
  journal = {Phys. Rev. Research},
  volume  = {2},
  pages   = {012032},
  year    = {2020},
  doi     = {10.1103/PhysRevResearch.2.012032}
}

@article{AuthierLoos2020JCP,
  author  = {Authier, Juliette and Loos, Pierre-Fran{\c{c}}ois},
  title   = {Dynamical kernels for optical excitations},
  journal = {J. Chem. Phys.},
  volume  = {153},
  pages   = {184105},
  year    = {2020},
  doi     = {10.1063/5.0028040}
}

@article{BaldwinKlaiber1947,
  title = {Photo-Fission in Heavy Elements},
  author = {Baldwin, G. C. and Klaiber, G. S.},
  journal = {Phys. Rev.},
  volume = {71},
  issue = {1},
  pages = {3--10},
  numpages = {0},
  year = {1947},
  month = {Jan},
  publisher = {American Physical Society},
  doi = {10.1103/PhysRev.71.3},
  url = {https://link.aps.org/doi/10.1103/PhysRev.71.3}
}

@article{LevingerBethe1950,
  title = {Dipole Transitions in the Nuclear Photo-Effect},
  author = {Levinger, J. S. and Bethe, H. A.},
  journal = {Phys. Rev.},
  volume = {78},
  issue = {2},
  pages = {115--129},
  numpages = {0},
  year = {1950},
  month = {Apr},
  publisher = {American Physical Society},
  doi = {10.1103/PhysRev.78.115},
  url = {https://link.aps.org/doi/10.1103/PhysRev.78.115}
}

\end{document}